\documentclass[journal=jacsat,manuscript=article]{achemso}

\usepackage{chemformula} 
\usepackage[T1]{fontenc} 
\usepackage{graphicx} 
\usepackage{tabularx, makecell, booktabs}
\usepackage{amssymb} 
\usepackage{xcolor}

\author{Grégoire Saerens }
\affiliation[ETH Zurich]
{Department of Physics, Institute for Quantum Electronics, Optical Nanomaterial Group, 8093 Zurich, Switzerland}

\author{Günter Ellrott}
\affiliation[FAU Erlangen]
{Department of Physics, Friedrich-Alexander-Universität Erlangen-Nürnberg (FAU), 91058 Erlangen, Germany}

\author{Olesia Pashina}
\affiliation[ITMO]
{School of Physics and Engineering, ITMO University, 191002 St. Petersburg, Russia}
\alsoaffiliation[Brescia]
{University of Brescia, Department of Information Engineering, Via Branze 38, 25123, Brescia, Italy}

\author{Ilya Deriy}
\affiliation[ITMO]
{School of Physics and Engineering, ITMO University, 191002 St. Petersburg, Russia}
\alsoaffiliation[HEU]
{Qingdao Innovation and Development Center, Harbin Engineering University, Sansha road 1777, Qingdao, 266000, Shandong, China}

\author{Vojislav Krsti\'{c}}
\affiliation[FAU Erlangen]
{Department of Physics,  Friedrich-Alexander-Universität Erlangen-Nürnberg (FAU), 91058 Erlangen, Germany}

\author{Mihail Petrov}
\affiliation[ITMO]
{School of Physics and Engineering, ITMO University, 191002 St. Petersburg, Russia}

\author{Maria Chekhova}
\affiliation[MPL Erlangen]
{Max-Planck Institute for the Science of Light, 91058 Erlangen, Germany}
\alsoaffiliation[FAU Erlangen]
{Department of Physics, Friedrich-Alexander-Universität Erlangen-Nürnberg (FAU), 91058 Erlangen, Germany}

\email{maria.chekhova@mpl.mpg.de}

\author{Rachel Grange}
\affiliation[ETH Zurich]
{Department of Physics, Institute for Quantum Electronics, Optical Nanomaterial Group, 8093 Zurich, Switzerland}
\email{grange@phys.ethz.ch}

\title{Second-Order Nonlinear Circular Dichroism in Square Lattice Array of Germanium Nanohelices}

\abbreviations{IR,SHG,SEM}

\begin{document}

\begin{abstract}

Second harmonic generation (SHG) is prohibited in centrosymmetric crystals such as silicon or germanium due to the presence of inversion symmetry.
However, the structuring of such materials makes it possible to break the inversion symmetry, thus achieving generation of second-harmonic.
Moreover, various symmetry properties of the resulting structure, such as chirality, also influence the SHG.
In this work we investigate second harmonic generation from an array of nanohelices made of germanium.
The intensity of the second harmonic displayed a remarkable enhancement of over 100 times compared to a non-structured Ge thin film, revealing the influence of interaction between nanohelices.
In particular, nonlinear circular dichroism, characterized through the SHG anisotropy factor $g_{SHG-CD}$, changed its sign not only with the helix handedness, but with its density as well.
We believe that our discoveries will open up new paths for the development of nonlinear photonics based on metamaterials and metasurfaces made of centrosymmetric materials.

     
\end{abstract}

\section{Introduction}
To feature second-order nonlinear effects, a material needs to have no inversion symmetry. But, even in materials without bulk second-order susceptibility $\chi^{(2)}$, the inversion symmetry can be lifted through structuring.
The simplest example is breaking of an inversion symmetry at the surface of a planar structure.
Second-order effects, such as second-harmonic generation (SHG), have been observed on the surfaces of metals and semiconductors already at the dawn of nonlinear optics~\cite{Bloembergen68}. The surface contribution becomes significant even for non-centrosymmetric materials if their bulk response is suppressed by the orientation of the crystal sample~\cite{Germer97} or in nanoscale structures where surface-to-volume ratio is relatively high~\cite{Wei17}. In plasmonic (metal) nanostructures, SHG is considerably enhanced if the structuring breaks the inversion symmetry at subwavelength scales. Plasmonic metasurfaces with non-centrosymmetric, also 2D-chiral, meta-atoms can reach SHG efficiencies of $\eta_{SHG} = P_{SHG}/P_{Fund}=10^{-9}$ at input powers $P_{Fund}$ of few $\mu$W~\cite{Krasnok18}, where $P_{SHG}$ is the measured SHG power. However, higher input powers are hardly tolerable because of the low damage threshold of plasmonic nanostructures. 

Dielectric metamaterials have the advantage of higher damage thresholds. Besides, unlike in metallic structures, where the electric field is nonzero only on the surface, dielectric materials allow the nonlinear response from the bulk as well. The most efficient second-order nonlinear effects are observed in metasurfaces made of materials with high bulk second-order susceptibility $\chi^{(2)}$, such as GaAs, AlGaAs, GaP, LiNbO$_3$ etc.~\cite{Li17, Krasnok18, Sain19, DeAngelis20, Weigand21}. Especially in the presence of geometric resonances, metasurfaces made of such materials enable SHG efficiency of up to $10^{-5}$ for pulsed pump~\cite{Liu16} or up to $10^{-7}$ for CW pump~\cite{Anthur20}.

Silicon (Si) and germanium (Ge) are commonly used all-dielectric materials that require techniques for second-order nonlinear applications, either enhancing the electric field at the surface~\cite{Makarov17,Wang22,Zhang19} or inducing externally an asymmetry in the crystal~\cite{Schriever10,Timurdogan17,Tan23}. However, these techniques are often limited to a specific material or a narrow bandwidth of operation wavelengths. For instance, a metasurface with asymmetric (T-shaped) silicon meta-atoms generated second harmonic with an efficiency of $10^{-9}$ under pumping with $3$ mW\cite{Liu19}. However, this effect was attributed solely to a resonance with a high ($10^5$) quality factor. 

\begin{figure}[h!]
\centering
\includegraphics{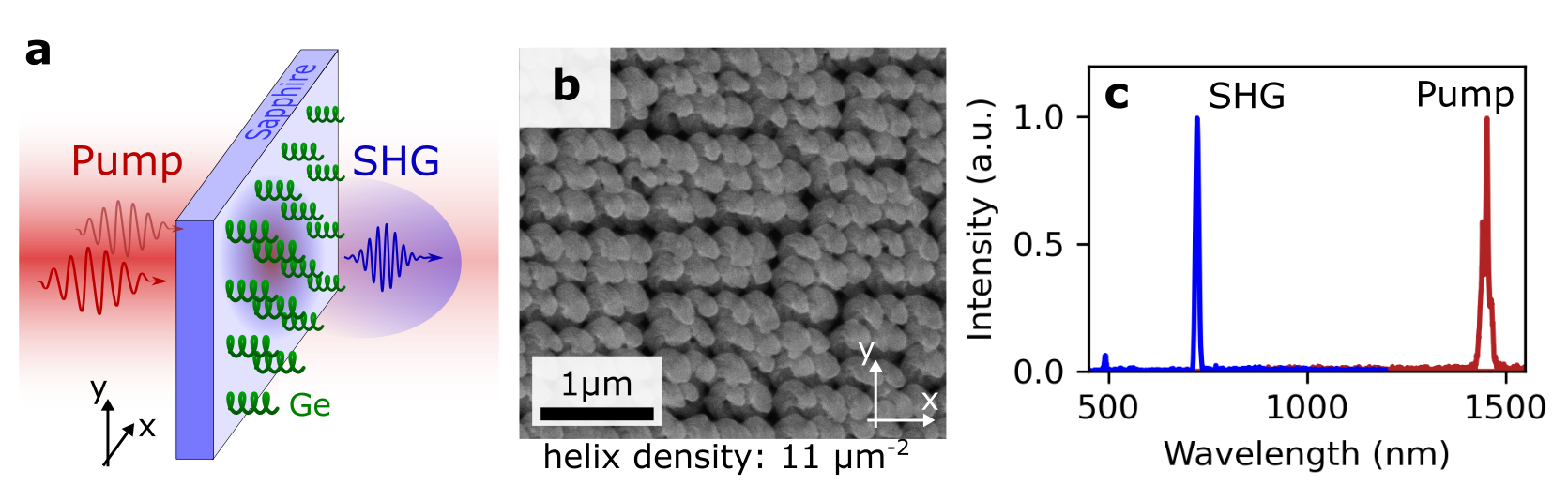}
\caption{(a) Schematic of Ge nanohelices arranged in a square lattice on a transparent sapphire substrate and generating second-harmonic light in the visible. (b) Top scanning electron microscopy (SEM) image of closely packed Ge nanohelices with a density of 11 nanohelices per $\mu$m$^2$. (c) Spectra of the pulsed laser light at 1450 nm wavelength (Pump) and of the filtered signal (SHG). The filtered intensity peak is exactly at half the wavelength of the pump, indicating SHG from the Ge nanohelices.}
\label{fig:1}
\end{figure}

In this paper, we structured a high-index semiconductor material into three-dimensional (3D) chiral elements\cite{Barron13}, breaking the central symmetry not on the atomic level, as is typical, but on the subwavelength level. We investigated SHG from Ge nanohelices, as shown schematically in Fig. \ref{fig:1}a, and demonstrated that the helical structure can yield a second-order nonlinear response exceeding the surface contribution by more than one order of magnitude. To support our experimental findings, we conducted simulations using the finite element method (FEM), through the full-wave electromagnetic modeling software COMSOL Multiphysics, comparing in particular SHG from nanohelices and from centrosymmetric tori. 

While circular dichroism is expected from chiral structures and has already been studied in the linear optical regime \cite{Ellrott23}, its manifestation in the SHG enables different applications - for instance, in nonlinear chiral sensing and imaging for material science and biology~\cite{Petralli93, Simpson02, Lee13}. Circular dichroism of the SHG is quantified through the SHG anisotropy factor $g_{SHG-CD}$, in line with established methodologies~\cite{Wang23}:

\begin{equation}
    g_{SHG-CD} = \frac{2(I_{lhcp}-I_{rhcp})}{(I_{lhcp}+I_{rhcp})}.
\end{equation}
where $I_{l/rhcp}$ is the intensity of the left- and right-handed circular polarized light, respectively. Centrosymmetric structures, such as Cu or Ge halides~\cite{Guo22,Wang23}, plasmonic nanoparticles~\cite{Huttunen11,Spreyer22}, dielectric dimers~\cite{Frizyuk21} or dielectric and plasmonic metasurfaces~\cite{Huttunen11, Mamonov12, Kim20}, have demonstrated circular dichroism of the SHG response in the visible and near-infrared region. Recently, there has also been a significant progress in theoretical description of SH dichroism in nanoantenna and metasurface structures\cite{Nikitina23, koshelev23}. Here we measure, for the first time, the SHG anisotropy in pure Ge, with a factor $g_{SHG-CD}$ that takes values up to 1 and changes sign around a density of 18 nanohelices per $\mu$m$^2$. This experimental observation is further supported by our numerical simulations, which predict a maximal value of $g_{SHG-CD}=1.7$ and the change of sign for similar nanohelix densities. We conclude by discussing the evidences of a collective effect, which was already observed in the linear optical regime\cite{Ellrott23}.

\section{Results and discussion}

We fabricated square-array arrangements of Ge nanohelices with subwavelength pitch and period, using the glancing angle deposition (GLAD) procedure as explained in our previous work~\cite{Ellrott23}. A top scanning electron microscopy image (SEM) is given in Fig. \ref{fig:1}b for the sample with a density of 11 nanohelices per $\mu m^2$. Here, we defined the two lattice axes of the square array as x- and y-axis. The fabrication method as well as additional images can be found in Section S1 of the Supporting Information. To investigate the impact of collective effects on the SHG, samples were grown with four different nanohelix densities, ranging from 44 to 5 helices per $\mu$m$^2$, corresponding to a spacing of 150 and 450 nm, respectively. For all the samples, the helical diameter is $D = 190$ nm and the helical pitch is $p_h = 360$ nm, with 2 turns, resulting in a total height of $720$ nm (see also Fig. S3 in the Supporting Information).

We performed the nonlinear measurements using a near-infrared tunable femtosecond laser as the excitation light source. The laser's power and polarization were controlled by motorized rotating half-waveplate retarders before and after a polarizing beam splitter, respectively. Switching between linearly- and circularly-polarized light could be achieved by adding and rotating a quarter-waveplate. The beam was then focused onto the center of an array (100 x 100 $\mu$m$^2$), resulting in a spot diameter of approximately 25 $\mu$m. The signal was collected using a 20x objective, filtered through two high-pass filters, and captured by either a CMOS camera or a spectrometer. Figure \ref{fig:1}c shows the spectra of the pump and the filtered signal, confirming the detection of just the SHG, namely at exactly half the excitation wavelength. Additional details regarding the setup and the procedure for the SHG power measurement can be found in Section S2 of the Supporting Information, together with further characterisation of the SHG for different wavelengths and powers.

\begin{figure}[h!]
\centering
\includegraphics{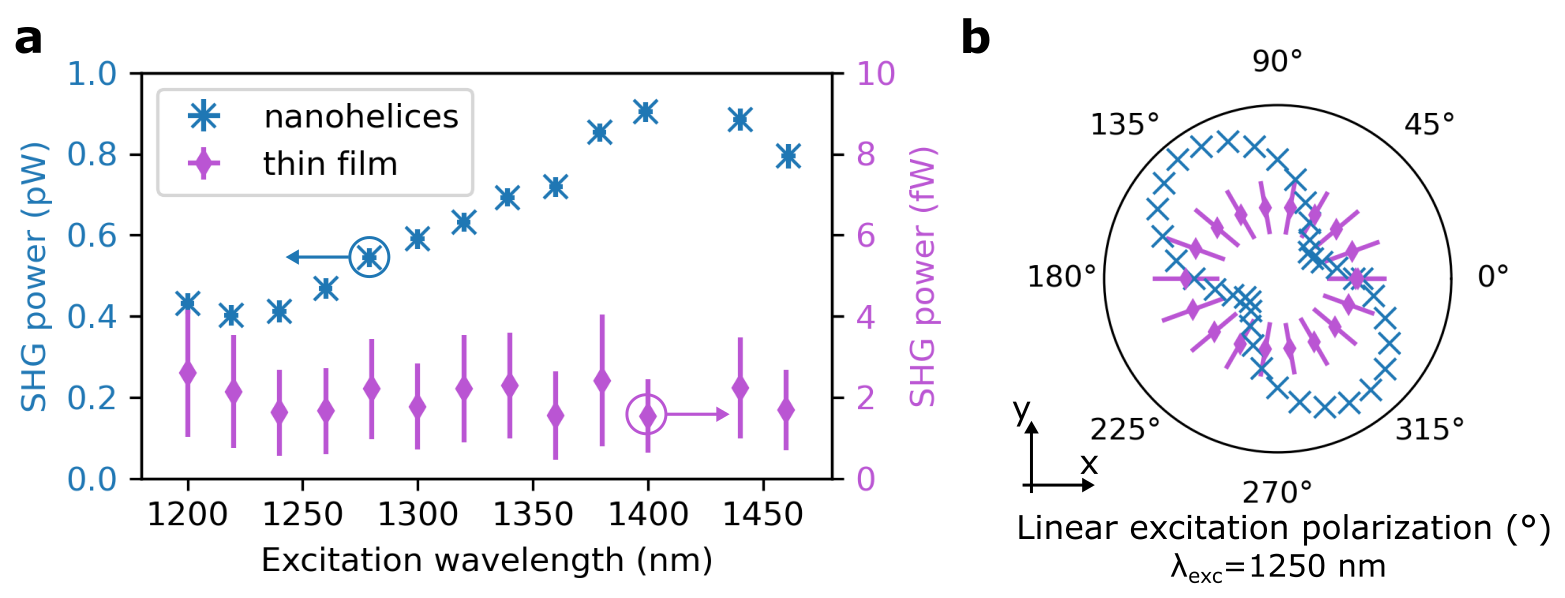}
\caption{(a) SHG power measured at different excitation wavelengths from closely packed Ge nanohelices (blue, 44 nanohelices per $\mu$m$^2$) and a comparison 50 nm thin Ge film (purple). The SHG intensity from the nanohelices is more than 100 times higher than that of a thin film. (b) Normalized SHG intensity for different excitation polarizations from the same nanohelix array as in (a) (blue crosses) and the Ge thin film (purple diamonds) at an excitation wavelength of 1250 nm.}
\label{fig:2}
\end{figure}

The SHG emission from Ge nanohelices under an excitation power of 100 mW was observed in transmission using the integration time of a few seconds. For further comparison, we measured the SHG intensity of a 50 nm thin film of Ge, as shown in Fig. \ref{fig:2}a. The thin film emitted a barely visible SHG signal, while closely packed Ge nanohelices demonstrated a substantial 100-fold SHG increase over the whole excitation range, from 1200 nm to 1400 nm. Meanwhile,  we calculated the surface increase from a thin film to Ge nanohelices (see Section S3 of the Supporting Information) and obtained only a factor of 12 for closely packed Ge nanohelices compared to a bare thin film. Similarly, we measured the SHG intensity for different densities of nanohelices and calculated the corresponding surface increase. In all cases, the surface increase of the nanohelix compared to a thin film was one order of magnitude smaller than the SHG increase. This lower increase indicates that SHG originates primarily from the central symmetry breaking through the sub-wavelength structuring, and not from the surface. To further differentiate the effect of the surface or structured SHG, we developed a simulation model to compute the SHG from an array of nanostructures. Our model is described in Section S4 and S5 of the Supporting Information and takes into account surface dipolar and bulk quadrupolar contributions to the second-order polarization in centrosymmetric materials~\cite{Makarov17,Falasconi01}. We conducted simulations of the Ge nanohelix array and compared it with a torus array, which is centrosymmetric and similar in volume, as well as with a thin film. The simulation results, given in Section S6 of the Supporting Information, indicate no circular dichroism from the torus array. Furthermore, the SHG intensity from a torus array is similar to the one for a planar structure. These calculations supported the experimental results, namely that the sub-wavelength structuring is the origin of the circular dichroism of the SHG and that the second-order nonlinear signal emerges from the non-centrosymmetry of the nanohelices. 

Even though the bandgap of Ge is around 1900 nm, SHG could still be observed in the near-infrared. The absorption coefficient at the excitation (SHG) wavelength of 1400 nm (700 nm) is around 1.1 $\mu$m$^{-1}$ (7.5 $\mu$m$^{-1}$) leading to 93$\%$ (40$\%$) of the light still transmitting through the 120 nm thick Ge, which is equivalent to two turns of the nanohelix. 

The SHG intensity was also measured for different excitation polarizations. The polar plots, which were obtained by rotating the second half-wave plate (without the quarter-wave plate added), reveal that nanohelices emitted stronger SHG for specific polarizations, while the Ge thin film produced an SHG signal independent of the excitation polarization, as shown in Fig. \ref{fig:2}b. The origin of the polarization dependence of the nanohelix could be found in the start and end point of the nanostructure, which restrained the possible material polarization directions. The calculated electric field at these points also showed higher intensities, as shown in the Section S7 of the Supporting Information.

\begin{figure}[h!]
\centering
\includegraphics{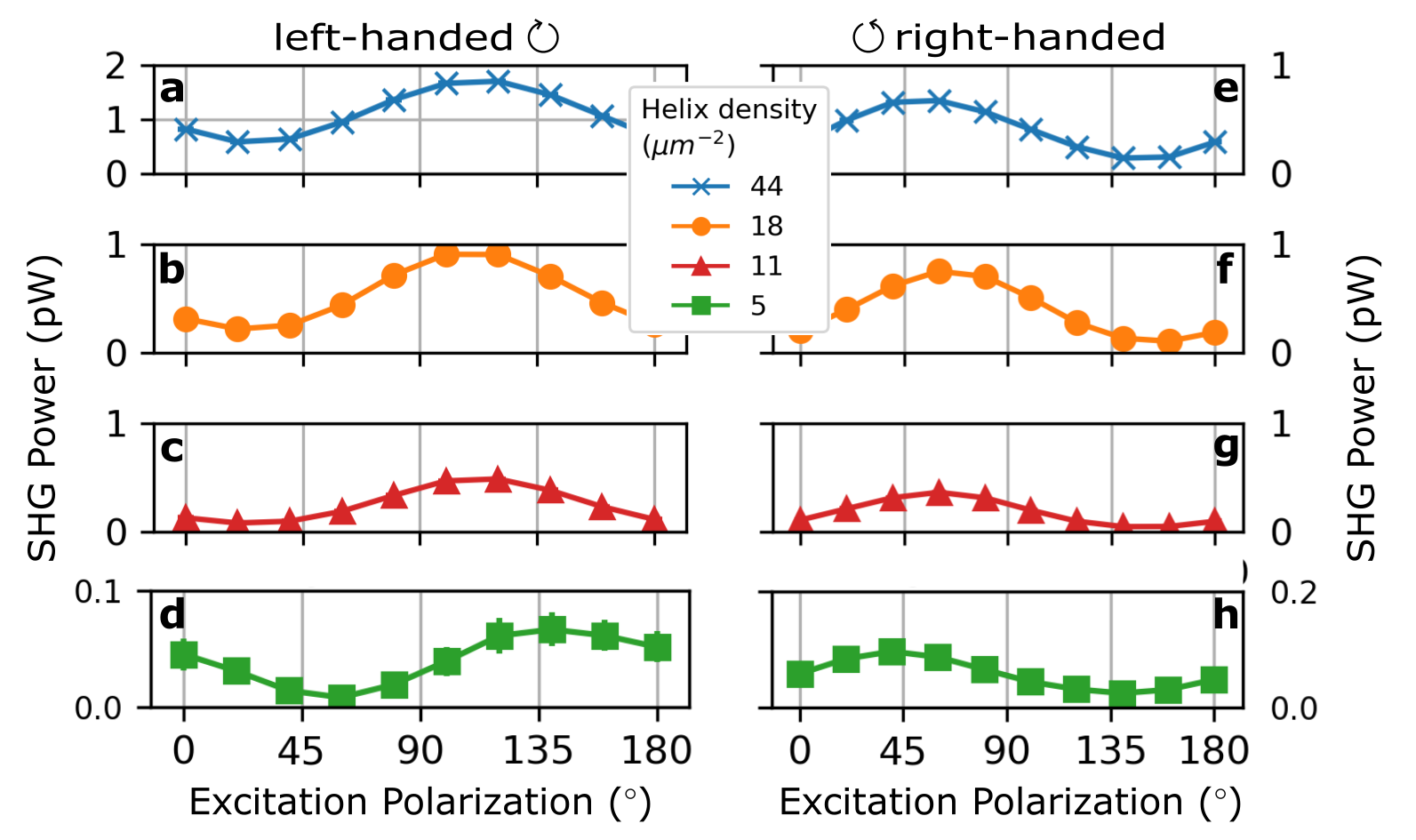}
\caption{Linear polarization orientation dependence of the SHG power from (a)-(d) left-handed and (e)-(h) right-handed Ge nanohelices at an excitation wavelength of 1450 nm. Results are shown from the highest to the lowest density of helices (top to bottom). The highest SHG efficiency is observed for a polarization around $120^{\circ}$ for left-handed nanohelices, relatively to the horizontal x-axis (see Fig. \ref{fig:1}), and around $60^{\circ}$ for right-handed nanohelices.}
\label{fig:3}
\end{figure}

We investigated further the linear polarization dependence of the SHG from Ge nanohelices with various densities and opposite handedness, as shown in Figure \ref{fig:3}. The SHG power was measured at a wavelength of 1450 nm, where the maximum signal was observed, and at an average power of 100 mW. Even for lower nanohelix densities, the SHG possessed a strong dependence on the orientation of the linear polarization, peaking at around $120^{\circ}$ ($60^{\circ}$) for left-handed (right-handed) nanohelices relatively to the horizontal x-axis (see Fig. \ref{fig:1}). This is similar to Fig. \ref{fig:2}b. The measured strong polarization dependence and the impact of the handedness were also observed in the simulations, as shown in Section S7 of the Supporting Information.

We defined the SHG anisotropy as $\rho_{SHG}=(I_{max}-I_{min}) /(I_{max}+I_{min})$, with $I_{max}$ and $I_{min}$ being the maximum and minimum SHG intensities, respectively,  measured for two different linear polarizations. In Table \ref{tbl:1}, we give the SHG anisotropy $\rho_{SHG}$, which ranges from 0.5-0.6 for closely packed nanohelices to 0.7-0.8 for a density of 11 nanohelices per $\mu$m$^2$.  These values of SHG anisotropy are smaller than the ones obtained numerically, as shown in the Section S7 of the Supporting Information. This lower value could be related to the quality of the nanohelix, as defects or imperfections could reduce the polarization dependency.

\begin{table}
  \caption{Measured SHG anisotropy ($\rho_{SHG}$) from the different Ge nanohelice samples.}
  \label{tbl:1}
  \begin{tabular}{cc}
    \hline
    Helix density ($\mu$m$^{-2}$)  & \makecell[c]{$\rho_{SHG}$\\ \hspace{2.5mm}left $\circlearrowright$ $|$ $\circlearrowleft$ right }  \\
    \hline
    44   & 0.49 | 0.65 \\
    18 & 0.61 | 0.75 \\
    11  & 0.71 | 0.77 \\
    5 & 0.77 | 0.60 \\
    \hline
  \end{tabular}
\end{table}

\begin{figure}[h!]
\centering
\includegraphics{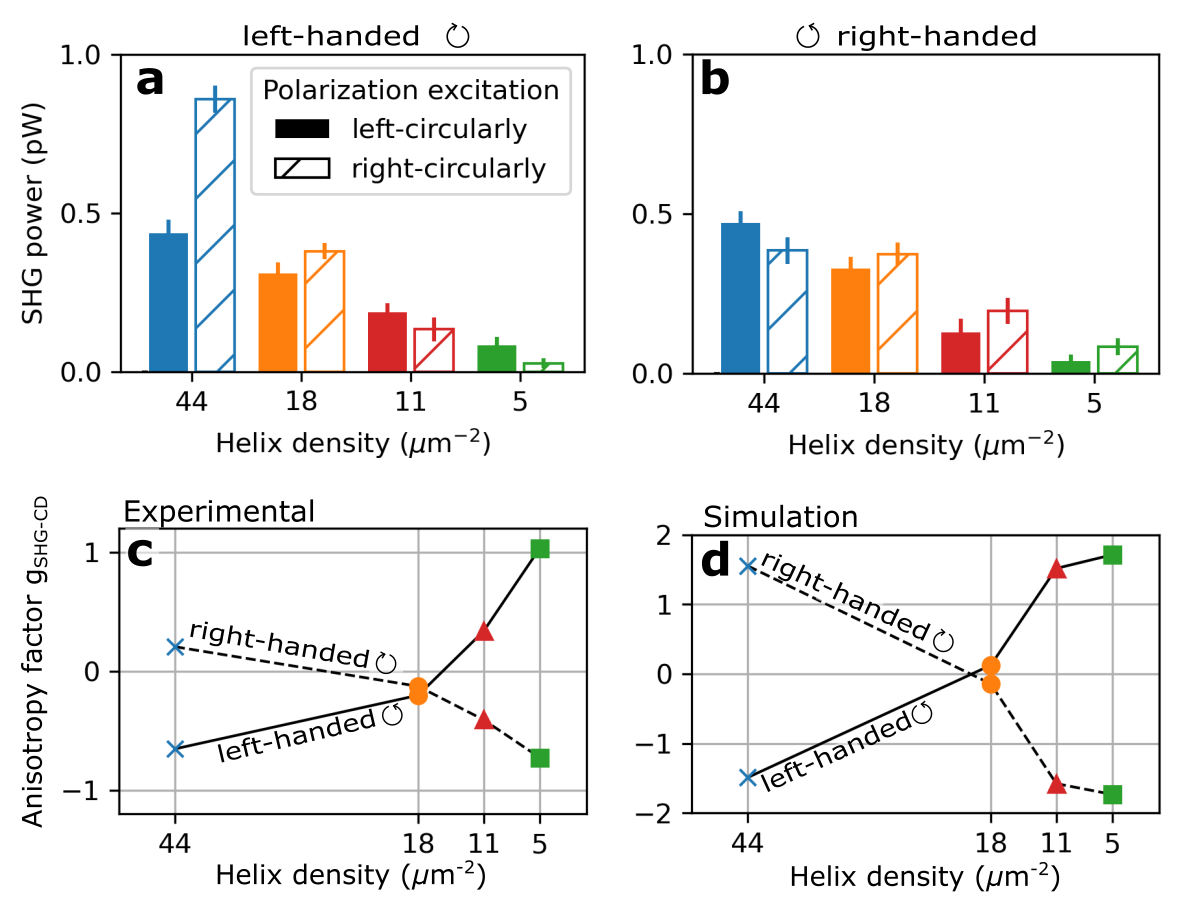}
\caption{SHG power from (a) left-handed and (b) right-handed Ge nanohelices arranged from the highest to the lowest density of nanostructures, progressing from left to right. Strong SHG efficiency is observed from closely packed left-handed helices under right-circularly polarized light and from closely packed right-handed helices under left-circularly polarized light. This nonlinear circular dichroism reverses with low-density nanohelices. (c,d) SHG anisotropy factor $g_{SHG-CD}$ measured (c) and simulated (d) for different densities and handednesses of the nanohelices.}
\label{fig:4}
\end{figure}

Similarly as in our previous study in the linear optical regime~\cite{Ellrott23}, we measured the circular dichroism of the SHG at 1450 nm excitation wavelength and 100 mW average power. We observed a strong nonlinear circular dichroism that switches with the nanohelix's handedness. The SHG from closely packed left-handed nanohelices was stronger for right-circularly polarized light (see Fig. \ref{fig:4}a), while from right-handed nanohelices, it was stronger for left-circularly polarized light (see Fig. \ref{fig:4}b). However, this circular polarization dependence reversed with an increase in the pitch. The SHG power from loosely packed left-handed nanohelices favored left-circularly polarized light, and conversely from loosely packed right-handed nanohelices. 

The SHG anisotropy factor $g_{SHG-CD}$ is illustrated in Fig. \ref{fig:4}c with different helix densities in the horizontal axis. Interestingly, we observed a zero crossing between 18 and 11 helices per $\mu$m$^{2}$. The values of $g_{SHG-CD}$ varied from 0.20 to -0.72 for right-handed nanohelices and from -0.65 to 1.0 for left-handed nanohelices. In comparison, the calculated $g_{SHG-CD}$ varied from $\pm$ 1.5 for the highest density of helices to $\mp$ 1.7 for the lowest, as shown in Fig. \ref{fig:4}d. A change of signs for nanohelix densities around 18 $\mu$m$^{2}$ was also visible. We suggest that this change is due to destructive interference arising at certain pitches, similarly to the change in circular dichroism that can be observed in plasmonic nanostructures when changing the density \cite{Valev09, Bautista18}. The discrepancy between the experimental and simulated results, especially for left- and right-handedness, could again be explained by the sample quality. The comparison of SHG from different samples grown with similar parameters is shown in Section S8 of the Supporting Information. From this, it is possible to estimate the uncertainty in the SHG measurement arising from fabrication. Nevertheless, simulations revealed a resonance close to the second-harmonic wavelength, which, with increasing nanohelix density, shifted towards the longer wavelength region of the spectrum. This is shown in Section S9 of the Supporting Information

\begin{figure}[h!]
\centering
\includegraphics{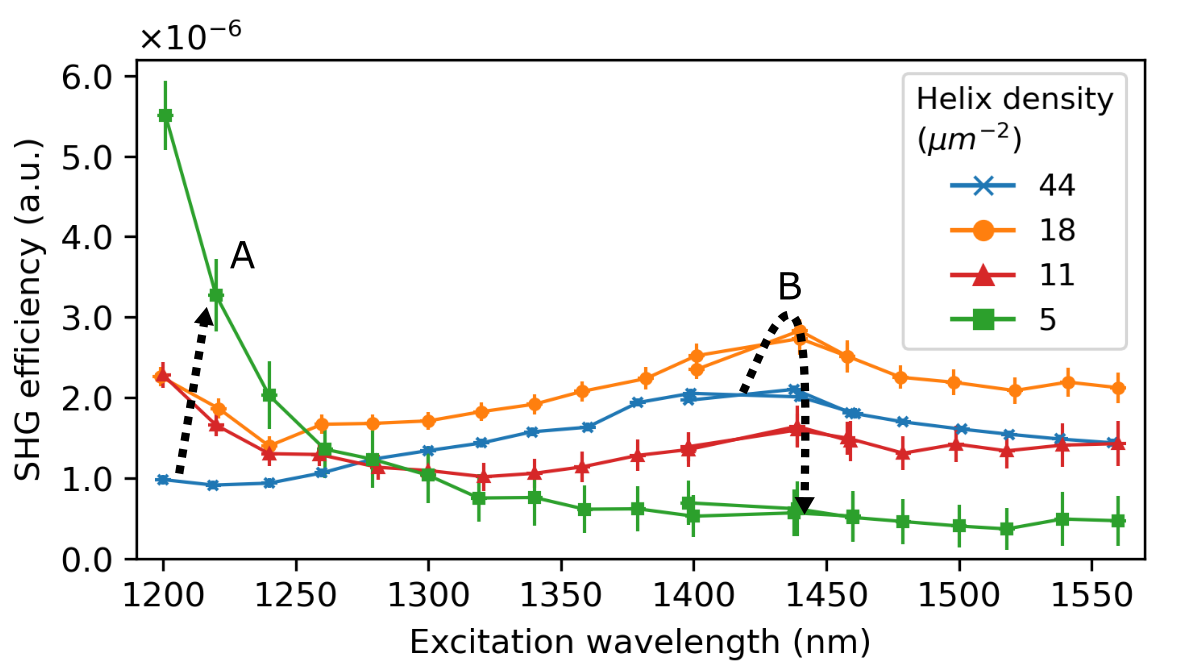}
\caption{SHG intensity spectrum from Ge nanohelices with different densities. Plotted as the SHG power normalized to the excitation power and the nanohelix density. For higher densities, the strongest SHG is observed around 1450 nm excitation, while for lower densities, the SHG intensity increases more drastically at shorter wavelengths and with decreasing density. Two resonances are observed (A,B) which appear for different densities.}
\label{fig:5}
\end{figure}

Similarly, we measured the SHG intensity for different excitation wavelengths and normalized them to the nanohelix density, as shown in Fig. \ref{fig:5}. The measurement was done in two steps, between 1200 nm and 1460 nm, and between 1400 nm and 1560 nm as it required a modified laser configuration at longer wavelengths (the pump laser wavelength was set to 820 nm instead of 800 nm). We added a correction factor of 1.45 for the second range, which took into account the new optical properties of the excitation laser, for instance the different pulse durations. On the one hand, the SHG dependence on polarization diminishes for all Ge nanohelices when exciting with shorter wavelength, as detailed in Section S10 of the Supporting Information. However, the SHG intensity per nanohelix density around 1200 nm excitation wavelength (resonance A in Fig. \ref{fig:5}) increased by a factor of 5 with larger spacing between nanohelices. We interpret this difference as an indication of a collective effect, where emission from densely packed Ge nanohelices interferes destructively. While the efficiency per nanohelix increases with decreasing nanostructure density, further exploration is needed - for example, by testing samples with different helical pitches or spacings, as well as measuring SHG below 1200 nm. On the other hand, a maximum SHG intensity per nanohelix density is observed for a density of 18 per $\mu$m$^2$. This second resonance (B) may arise from constructive interference at a specific pitch (see also comparison in Section S11 of the Supporting Information). These insights hold potential for developing efficient arrays of nonlinear emitters from a wide range of materials, not being limited by the bulk $\chi^{(2)}$ tensor. Possible designs based on rescaling the structure should also give control over a broad range of excitation wavelengths.

\subsection{Conclusion}

In summary, we fabricated periodic nanohelices arrays from Ge, a centrosymmetric material, and successfully measured SHG from it. The measured intensity was greatly enhanced compared to a Ge thin film, by a factor of more than 100. This two-order of magnitude enhancement could not be explained by the increase of surface only, which would increase only by about 12 times for the closest packed helices. We suggested a collective effect that was further confirmed by the observations of SHG with different linearly and especially circularly-polarized light. We measured the circular dichroism with the SHG anisotropy factor $g_{SHG-CD}$ and obtained values of up to 1 depending on the helix density. This SHG anisotropy factor changed sign for helices grown with different handedness, as expected, but also changed sign with nanohelix densities around 15 $\mu$m$^{-2}$. Finally, we measured the SHG spectrum from 1200 nm to 1560 nm and compared the SHG efficiency normalized to the nanohelix density, and observed at shorter wavelengths a much higher nonlinear signal for lower nanohelix densities, confirming the role of collective effects in shaping their optical response. Efficient arrays of nonlinear emitters, incorporating truly chiral elements like nanohelices or spiral pyramids, can be structured in widely utilized centrosymmetric materials, such as Si or Ge. Our results present new opportunities for second-order nonlinear optical devices with strong optical activity for applications in all-optical signal processing, spectroscopy or quantum optics.  

\subsection{References}

\begin{acknowledgement}

The authors thank Vitaliy Sultanov and Ülle-Linda Talts for helpful discussions. This work was supported by the European Union’s Horizon 2020 research and innovation program from the European Research Council under the Grant Agreement No. 714837 (Chi2-nano-oxides). The authors achnowledge the support by the Deutsche Forschungsgemeinschaft (DFG, German Research Foundation), project ID 429529648 TRR 306 QuCoLiMa (“Quantum Cooperativity of Light and Matter”). The numerical simulations were performed within the Russian Science Foundation project (no. 22-12-00204).

\end{acknowledgement}

\begin{suppinfo}

The following files are available free of charge.
\begin{itemize}
  \item Supporting Information
\end{itemize}

\end{suppinfo}


\bibliography{00_Bib_Collection}

\end{document}